\documentclass{sf2a-conf2012}
\usepackage{graphicx}
\usepackage{hyperref}
\usepackage[]{natbib}  
\usepackage[cyr]{aeguill}
\usepackage{epstopdf}

\def\BibTeX{{\rm B\kern-.05em{\sc i\kern-.025em b}\kern-.08em
    T\kern-.1667em\lower.7ex\hbox{E}\kern-.125emX}}
\bibpunct{(}{)}{;}{a}{}{,}  %%%%%%%%%%%%%  A&A bibliography style
\begin{document}

\TitreGlobal{SF2A 2012}

%%-----------------------------------------------------------------
%%      the top matter
%%

\title{First LOFAR results on galaxy clusters}

\runningtitle{First LOFAR results on galaxy clusters}

\author{C. Ferrari}\address{Laboratoire Lagrange, UMR 7293, Universit\'e de Nice Sophia-Antipolis, CNRS, Observatoire de la C\^ote d'Azur, 06300 Nice, France }

\author{I. van Bemmel}\address{ASTRON, Postbus 2, 7990 AA Dwingeloo, The Netherlands}

\author{A. Bonafede$^{3,}$}\address{Hamburger Sternwarte, Universit\"at Hamburg, Gojenbergsweg 112 21029 Hamburg, Germany}\address{Jacobs University Bremen, Campus Ring 1, 28759, Bremen, Germany}

\author{L. B\^irzan}\address{Leiden Observatory, Leiden University, PO Box 9513, 2300 RA, Leiden, The Netherlands}

\author{M. Br\"uggen$^{3,}$$^{4}$}

\author{G. Brunetti}\address{INAF/Istituto di Radioastronomia, via Gobetti 101, 40129 Bologna, Italy}

\author{R. Cassano$^6$}

\author{J. Conway}\address{Onsala Space Observatory, Dept. of Earth and Space Sciences, Chalmers University of Technology, 43992 Onsala, Sweden }

\author{F. De Gasperin$^{3}$}

\author{G. Heald$^2$}

\author{N. Jackson}\address{Jodrell Bank Center for Astrophysics, School of Physics and Astronomy, The University of Manchester, Manchester M13 9PL, UK }

\author{G. Macario$^1$}

\author{J. McKean$^2$}

\author{A. R. Offringa}\address{RSAA, The ANU Mt Stromlo Observatory, Australia}

\author{E. Orr\`u$^{2,}$}\address{Department of Astrophysics/IMAPP, Radboud University Nijmegen, PO Box 9010, 6500 GL Nijmegen, The Netherlands }

\author{R. Pizzo$^{2}$}

\author{D. A. Rafferty$^{5}$}

\author{H. J. A. R\"ottgering$^{5}$}

\author{A. Shulevski}\address{Kapteyn Astronomical Institute, PO Box 800, 9700 AV Groningen, The Netherlands}

\author{C. Tasse}\address{GEPI, Observatoire de Paris-Meudon, 5 place Jules Janssen, 92190, Meudon, France}

\author{S. van der Tol$^{5}$}

\author{R. J. van Weeren$^{5,2,}$}\address{Harvard-Smithsonian Center for Astrophysics, 60 Garden Street, Cambridge, MA 02138, USA}

\author{M. Wise$^{2,}$}\address{Astronomical Institute ``Anton Pannekoek'', University of Amsterdam, Postbus 94249, 1090 GE Amsterdam, The Netherlands}

\author{J. E. van Zwieten$^2$, on behalf of the LOFAR collaboration}

%% Keep this line, even if the page will be settled afterwards.
\setcounter{page}{237}

%%-----------------------------------------------------------------

\maketitle

%%-----------------------------------------------------------------
%%        The abstract
%% 
%%  Warning!  within the abstract:
%%  - do not use macros. 
%%  - do not use commands like: \cite, \citet, \citep ... etc.

\begin{abstract}
Deep radio observations of galaxy clusters have revealed the existence of diffuse radio sources related to the presence of relativistic electrons and weak magnetic fields in the intracluster volume. The role played by this non-thermal intracluster component on the thermodynamical evolution of galaxy clusters is debated, with important implications for cosmological and astrophysical studies of the largest gravitationally bound structures of the Universe.

The low surface brightness and steep spectra of diffuse cluster radio sources make them more easily detectable at low-frequencies. LOFAR is the first instrument able to detect diffuse radio emission in hundreds of massive galaxy clusters up to their formation epoch ($z \sim$ 1). We present the first observations of clusters imaged by LOFAR and the huge perspectives opened by this instrument for non-thermal cluster studies.
\end{abstract}

%% Insert the keywords (to appear in the ADS indexing)
%% Keywords must be separated by a comma

%%-----------------------------------------------------------------

\section{Introduction}

Galaxy clusters are both powerful cosmological tools and unique astrophysical laboratories to study the evolution and interaction processes of baryons along the history of the universe \citep[see e.g.][and references therein]{Voit05}. Number counts of clusters as a function of mass and redshift can give important constraints on cosmological parameters \citep[e.g.][]{Borgani01,Vikhlinin09}. We firstly need to be able to identify clusters -- through the multi-wavelength emission related to their different components or through the gravitational lensing effect -- and to measure their redshift. And of course we need to be able to estimate their mass from observable quantities (e.g. X-ray luminosity; optical richness; velocity dispersion of cluster members; temperature and density profiles of the thermal intracluster medium; ... ). For this, a detailed characterization of the complex gravitational and non-gravitational physical processes acting on galaxy clusters is essential \citep[e.g.][]{Bohringer10}.

\begin{figure}[ht!]
 \centering
 \includegraphics[width=1\textwidth,clip]{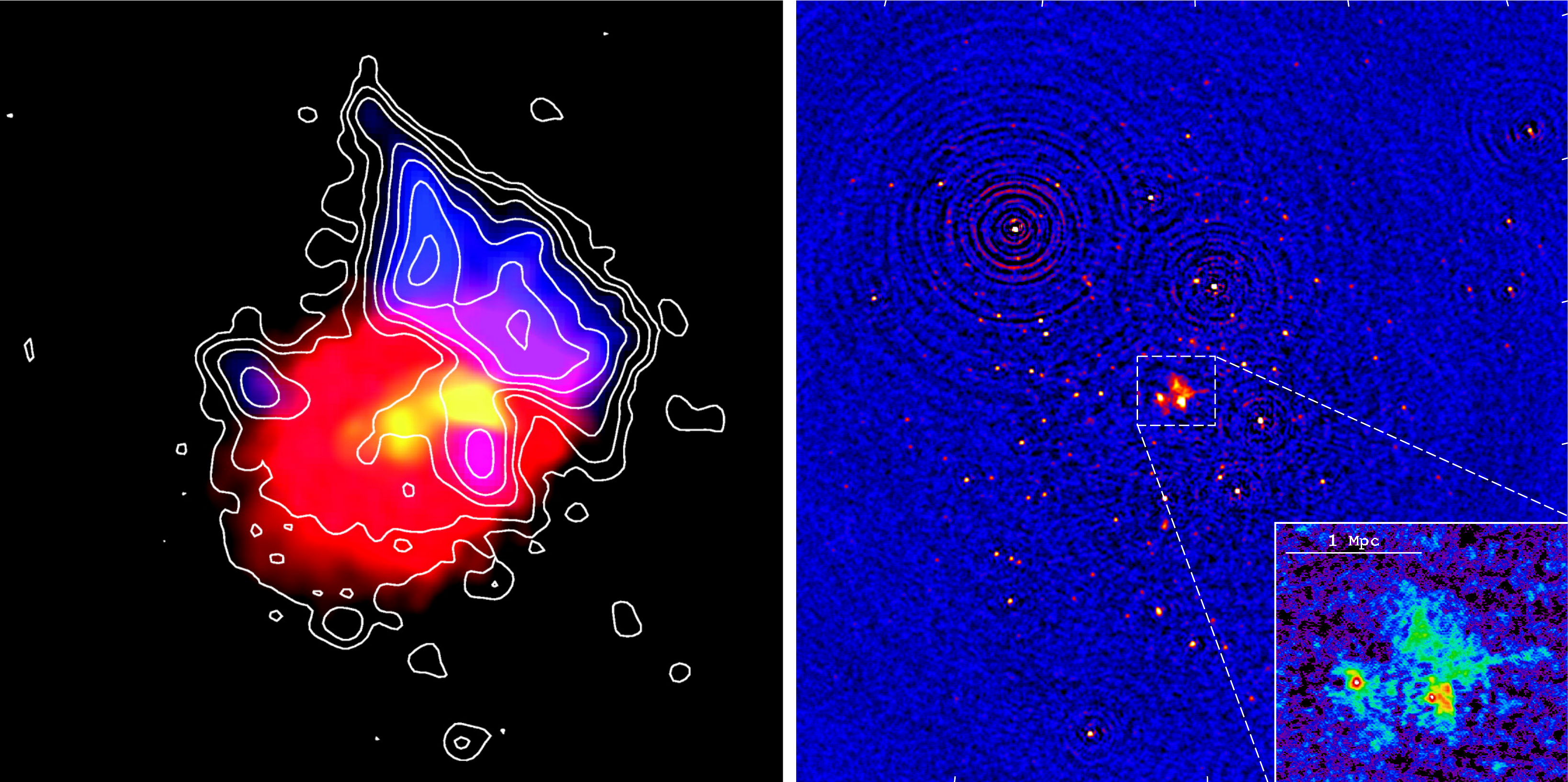}  
  %% Note the ABSENCE of the extension .pdf , .eps or .ps  !
  \caption{Radio observations of A\,2256 at $\approx$1400 MHz (white contours in the left panel; VLA observations) and at $\approx$ 60 MHz (color maps on the right; LOFAR observations). On the left, radio contours are overlaid on the Chandra X-ray image of A\,2256. The right image shows the low-resolution ($80'' \times 80''$) large-scale LOFAR map of the cluster, with a zoom in its central field shown at higher resolution ($22 '' \times 26 ''$) in the inset. Figures are extracted from \citet{Clarke06} and \citet{vanWeeren12}. }
  \label{ferrari:fig1}
\end{figure}

In the last years, much interest has been paid to the influence of the non-thermal component of the intracluster medium on the thermo-dynamical evolution, heating transport processes and mass determination of galaxy clusters \citep[e.g.][]{Parrish09,Lagana10}. The existence of relativistic electrons and magnetic fields in the volume in between cluster galaxies has been pointed out by radio observations of diffuse synchrotron sources \citep[see e.g.][for a recent review]{Feretti12}. Mpc-scale cluster radio sources are generally divided in ``halos'' and ``relics'' depending on their position in the cluster, morphology, polarization properties \citep{Ferrari08}. The left panel of Fig.\,\ref{ferrari:fig1} shows the radio contours overlaid on the X-ray map of one of the most studied nearby galaxy clusters, Abell\,2256 (A\,2256 in the following). This system hosts both a bright and elongated radio relic in its North-West external region and a central, very low-surface brightness radio halo. The morphology of this latter source is very similar to the X-ray emission of the cluster, which is related to the thermal intracluster component. 

A detailed understanding of the origin of the intracluster non-thermal component is still missing \citep[e.g.][]{Dolag08,Brunetti11}. Theoretical models need to be compared to statistical samples of clusters emitting at radio wavelengths, while only a few tens of radio relics and halos are known up to now \citep{Feretti12,Nuza12}. Diffuse intracluster radio sources are generally characterized by steep synchrotron spectra ($\alpha \geq -1$, $S_{\nu} \propto \nu^{\alpha}$). This, together with their low-surface brightness and the possible spectral steepening at high radio frequencies due to electron aging, make them difficult to be imaged in the GHz regime and more easily detectable at the long wavelengths observed by LOFAR. This instrument is therefore expected to be the first to allow the detection of diffuse radio emission in hundreds of massive galaxy clusters up to z $\sim$ 1 \citep{Cassano10}.

%%---------------------
    
\section{First LOFAR observations of galaxy clusters}
%%---------------------------------

Thanks to its large field of view (FoV), frequency coverage and high sensitivity, LOFAR is an excellent survey instrument that opens enormous perspectives for the study of radio plasma in the ICM \citep{Rottgering11}. 

Issues related to direction dependent effects \citep[such as beam variations in both time and frequency, or ionospheric distortions of the wavefront propagation on scales smaller than the LOFAR FoV; see e.g.][]{Tasse12} are particularly critical in the calibration and imaging phases of LOFAR data. For this, a detailed and complex data reduction chain is being tested and implemented during the LOFAR commissioning phase \citep{Heald11}. Very briefly, after the initial phase of data flagging and compression, the brightest radio sources in the sky (the so-called ``A-team''), that during observations move in and out the side-lobes of the station beams, need to be subtracted. New calibration and imaging algorithms, taking into account direction dependent effects as well as the non-coplanarity of the array, are under development \citep{Tasse12}.

LOFAR project started his commissioning activities about three years ago. A lot of progress has been made thanks to commissioning events called ``Busy Weeks'', where expert commissioners get together and work on specific topics. Besides the ``Busy Weeks'', since September 2010 we kept the commissioning work very active with the ``Busy Days'', taking place every other week at ASTRON, with EVO sessions organized for commissioners located all around the world.

\begin{figure}[ht!]
 \centering
 \includegraphics[height=0.4\textwidth,clip]{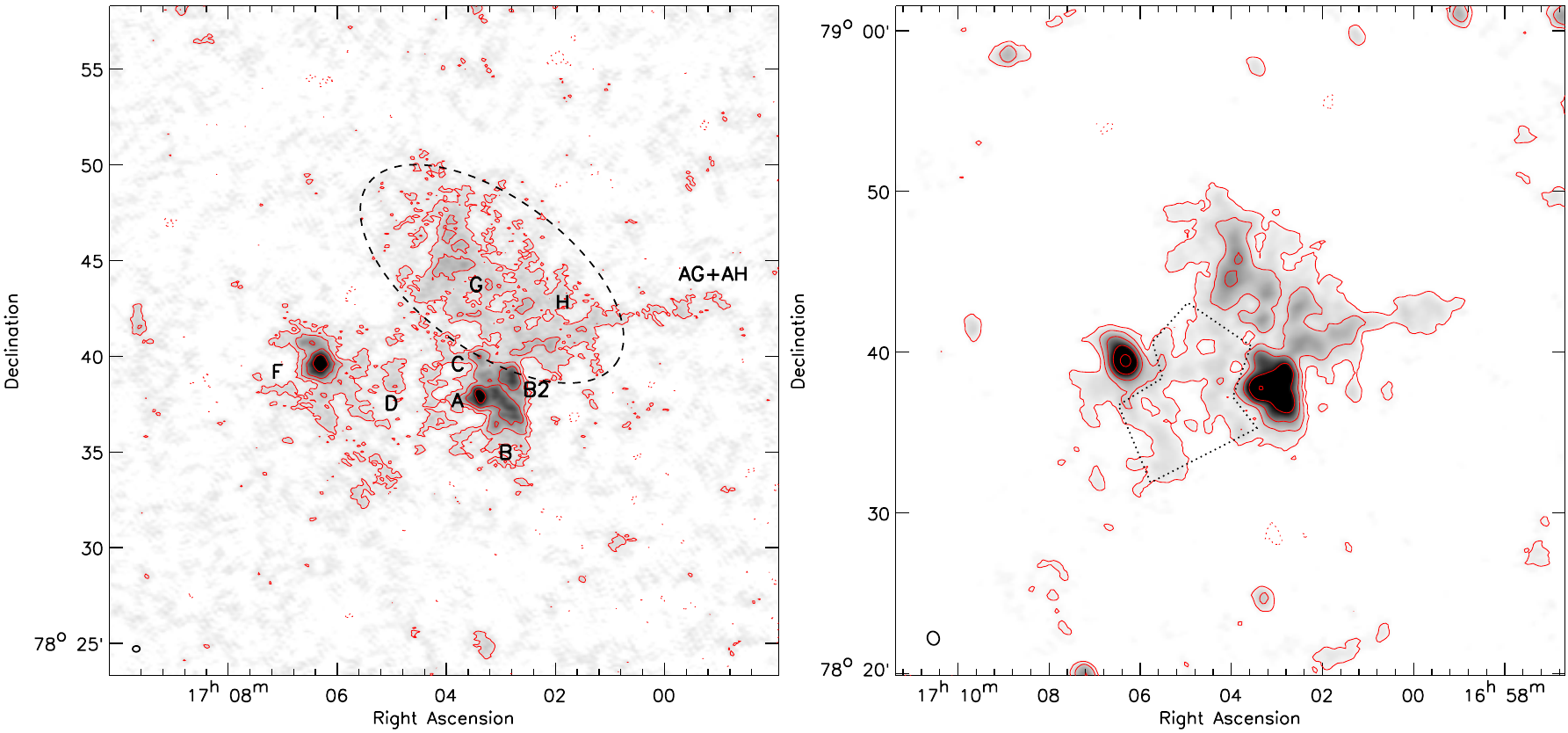}
 \includegraphics[height=0.4\textwidth,clip]{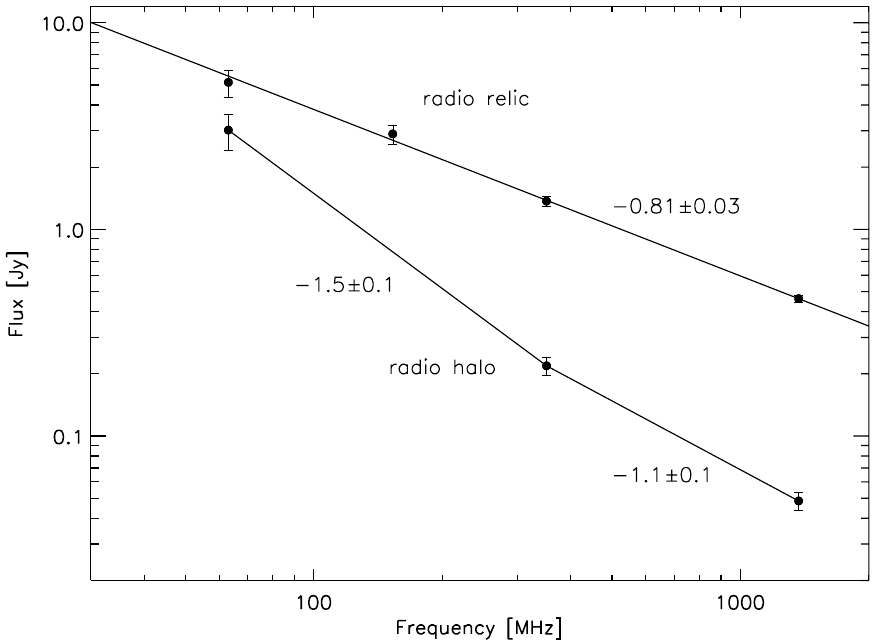}      
%% Note the ABSENCE of the extension .pdf , .eps or .ps  !
  \caption{{\em Top:} high- and low-resolution ($22 '' \times 26 ''$, left, and $52 '' \times 62 ''$, right) LOFAR maps of the galaxy cluster A\,2256 at $\approx$ 60 MHz. The radio relic is indicated by a dashed ellipse on the left, while the area used to measure the radio halo spectral index is shown by the dotted polygon on the right. {\em Bottom:} radio spectra for the relic and the halo in A\,2256. Figures extracted from \citet{vanWeeren12}.}
  \label{ferrari:fig2}
\end{figure}

One of the very first galaxy clusters observed by LOFAR has been A\,2256 \citep{vanWeeren12}. X-ray and optical observations provide strong evidence that A\,2256 is undergoing a merger event between a main cluster, a major sub-structure and, possibly, a third infalling group. As introduced in the previous section and shown in Fig.\,\ref{ferrari:fig1}, it is a nearby ($z$ = 0.0581) system that contains a giant radio halo, a relic and a large number of tailed radio galaxies. The relic has a large integrated flux compared to other sources of the same class, of about 0.5 Jy at 1.4 GHz. A spectral analysis by \citep{Brentjens08} shows that the radio halo component dominates the integrated cluster spectrum at very low frequencies. The large angular extent of the diffuse emission and its large integrated flux make A\,2256 a prime target for low-frequency observations which typically suffer from low spatial resolution and sensitivity, compared to observations at high frequencies. 

The LOFAR 63 MHz image reveals some of the well known tailed radio sources, the main relic, and part of the radio halo (Fig.\,\ref{ferrari:fig2}), which is more easily detectable on the lower resolution map. The integrated fluxes of the radio halo and relic are difficult to be measured because they are partly blended with some of the complex head-tail radio sources in the cluster. To estimate their flux contribution, both the high and low resolution images have therefore been used \citep[see][for more details]{vanWeeren12}. 

In the case of radio relics, the origin of cosmic ray electrons is generally explained in the framework of the diffusive shock acceleration (DSA) theory. The spectral index of A\,2256 radio relic, integrated over the full extent of the source, is however too flat ($\alpha = 0.81 \pm 0.03$) to be explained by classical DSA. Different hypotheses have been proposed, such as the fact that we are observing a relic related to a young shock, in which energy injection and losses are not yet balanced, or that we are observing electron re-acceleration in a possibly inhomogeneous shock downstream region, or finally that our flux measurements are severely affected by projection effects \citep{vanWeeren12}. 

Also the radio halo spectral shape is puzzling, since it presents a somehow unexpected low-frequency steepening (see Fig.\,\ref{ferrari:fig2}). This could be related to: a) the superposition of two (or more) spectral components \citep[see also][]{Kale10}, b) observation of turbulent re-acceleration -- giving rise to the steep spectrum -- plus hadronic component -- related to the flat part of the spectrum, or c) inhomogeneous turbulent re-acceleration, whose efficiency changes with space and/or time in the emitting volume \citep{vanWeeren12}.

Several other clusters have been observed during the LOFAR commissioning phase. Preliminary results have been presented at the meeting ``LOFAR's view of galaxy clusters'' held last spring in Nice and can be found at the conference web page (see \url{https://gandc.oca.eu/spip.php?article398}). The emission from halos, relics and head-tail radio galaxies observed at higher frequencies is generally fully recovered in all the observed clusters, with in addition evidence of more extended low-frequency radio emission (de Gasperin et al. submitted, Pizzo et al. in prep., Macario et al. in prep., Bonafede et al. in prep., Orr\`u et al. in prep.). We can conclude that LOFAR commissioning observations of galaxy clusters have started to show the great potential of this instrument for the study of the non-thermal component in large-scale structures.

% Optional acknowledgements
% -------------------------
\begin{acknowledgements}
Chiara Ferrari and Giulia Macario acknowledge financial support by the {\it ``Agence Nationale de la Recherche''} through grant ANR-09-JCJC-0001-01. 
\end{acknowledgements}

%% The following lines are required when using BibTEX (strongly encouraged!):
\bibliographystyle{aa}  % A&A bibliography style file (aa.bst)
\bibliography{ferrari} % your references in file: Yourfile.bib

\end{document}